\documentclass{aa}
\def \xmm {XMM-Newton}
 
\def \nh {N${\rm _H}$}

\def \arcmin {\hbox{$^\prime$}}
\def \arcsec {\hbox{$^{\prime\prime}$}}

\def \rchisq {$\chi_{\nu} ^{2}$}
\def\approxgt{\mathrel{\hbox{\rlap{\lower.55ex \hbox {$\sim$}}
        \kern-.3em \raise.4ex \hbox{$>$}}}}
\def\approxlt{\mathrel{\hbox{\rlap{\lower.55ex \hbox {$\sim$}}
        \kern-.3em \raise.4ex \hbox{$<$}}}}
\usepackage{graphicx,float}
\begin{document}
\title{X-ray background  measurements with XMM-Newton EPIC }
\author{D. H. Lumb \inst{1}
	\and  R. S. Warwick \inst{2}
	\and  M. Page  \inst{3}
     \and A. De Luca\inst{4}$^{,}$\inst{5} 
\thanks{This work is based on observations made with the XMM-Newton, an
ESA science mission with instruments and contributions directly funded by
ESA member states and the USA (NASA).} 
}

\institute{Science Payload Technology Divn., Research and Science Support Dept. of ESA, ESTEC, 2200 AG Noordwijk, Netherlands \and X-ray Astronomy Group, University of Leicester, Leicester LE1 7RH, England \and Mullard Space Science Lab, University College London, Holmbury St Mary, England \and Istituto di Fisica Cosmica ``G Occhialini'',CNR, via Bassini 15, I-20133 Milano Italy \and Universita' di Milano Bicocca, Dipartimento di Fisica, Piazza della Scienza 3, I-20126 Milano Italy}

\offprints{D. Lumb (dlumb@rssd.esa.int)}

\date{Received date  / Accepted date}  

\titlerunning{XMM EPIC in-orbit background} 
\abstract{ 
We discuss the methods used to  compile a high signal-to-noise 
dataset representative of both the instrumental and cosmic background 
signal measured at high galactic latitude by the XMM-Newton EPIC cameras.  
The characteristics of the EPIC background  are described and the potential 
applications of the derived dataset in  general science analysis are 
outlined.  In the case of the cosmic X-ray background, the transition  
between  a hard power-law spectrum (due to the integrated emission  
of unresolved, largely  extragalactic, point  sources) and a softer 
thermal spectrum (produced by hot plasma associated with the Galactic 
plane and halo) is unambiguously detected around $\sim 1$ keV. We derive 
a value for the intensity of the power-law component of $2.15 \pm 0.26$ 
10$^{-11}$ erg cm$^{-2}$ s$^{-1}$ deg$^{-2}$ in the 2--10 keV band (Normalisation
at 1keV of 11.1 photons cm$^{-2}$s$^{-1}$sr$^{-1}$ keV$^{-1}$). 
The implication is that recent, very deep {\it Chandra} observations have 
resolved  $\sim$ 70 -- 90\% of the 2--10 keV background into discrete 
sources. Our measurement is towards the higher end of the range of quoted background
normalisations.

}

\maketitle

\section{Introduction}
The  instrumental  backgrounds   experienced  in  both  the Advanced CCD
Imaging Spectrometer  (ACIS) on {\it  Chandra}  (Weisskopf et al. \cite{Weisskopf}) 
and the European Photon Imaging  Camera (EPIC) on \xmm~ are rather similar, as
expected from their comparable orbits and detector technologies. These
comprise of  a combination  of the  unrejected components  of  direct and
Compton  scattered cosmic  rays, and  the flux  of particles  from the
magnetosphere  which  are  focused  by  the  mirror  system  onto  the
detector. Especially  for \xmm, whose larger effective  area and field
of view offer  some advantages for such observations,  the analysis of
very extended  and faint objects  (clusters of galaxies,  for example)
may  be  frustrated by  the  inability  to  account properly  for  the
instrumental background signal.

A thorough understanding of  this instrumental background is important
for analysing the  true Cosmic X-ray Background (CXB).  For many years
the interpretation of the hard CXB as the integrated light of faint
extragalactic X-ray sources, mostly Active Galactic Nuclei (AGN),
has been hampered by the paradox that the
diffuse  spectrum  at energies  $\approxgt$~1~keV  did  not match  the
spectral form  of the major AGN populations.   Much of the
{\em  soft} diffuse  X-ray background  was resolved  by ROSAT  (in the
Lockman Hole for example, Hasinger et al \cite{Hasinger98}), where the
majority   of   individual   sources   have indeed   been   identified  
with AGNs. However the ROSAT energy band only marginally
overlapped with the harder ($\geq$ 2 keV) band in which the spectrum
of the CXB has been best determined by non-imaging missions such
as HEAO-1 A2  ( Garmire et al. \cite{Garmire}, Marshall et al. \cite{Marshall}).

Currently the characterisation of the faintest X-ray source populations
in the 2-10 keV band  is being  revolutionised by observations
made by the  {\it Chandra} and  XMM-Newton Observatories.   Their
capabilities
are  somewhat complementary: the  unprecedented {\it  Chandra} angular
resolution  (Van  Speybroek et  al.   \cite{vanSpeybroek}) allows  for
negligible background and ultimate source detection sensitivity; while
the  XMM-Newton telescopes  (Jansen et  al.  \cite{Jansen})  offer the
largest ever focused area for unmatched photon gathering power.

The  first deep  field observations  performed by  these observatories
(Giacconi  et al  \cite{Giacconi}; Hasinger  et  al.  \cite{Hasinger};
Hornschemeier   et  al.    \cite{Hornschemeier};   Mushotzky  et   al.
\cite{Mushotzky})  have   confirmed  these  promises.   {\it  Chandra}
observations   to   a    source   limiting   sensitivity   of   $\sim$
2 10$^{-16}$  erg   cm$^{-2}$  s$^{-1}$  (0.5--2~keV)  resolved
$\sim$80\% of  the background,  and found many  hard spectra  at faint
levels, thus providing the solution of the  ``spectral  paradox'' (namely
the difference between the spectrum of  the background and the spectrum of
bright AGN).   Hornschemeier et al.  \cite{Hornschemeier}  also note an
increase  in the proportion of  normal galaxies  at flux  levels $\leq$3
10$^{-16}$  erg  cm$^{-2}$ s$^{-1}$.   XMM-Newton  pushed the  limits
further  than {\it  Chandra} in  the 5--10~keV  band, reaching  2.4 
10$^{-15}$    erg    cm$^{-2}$    s$^{-1}$    (Hasinger    et    al.
\cite{Hasinger}).  A definitive statement  about the {\em fraction} of
background  resolved  depends  upon  solving a  long-standing  problem
concerning  a discrepancy  of 30\%  between different  measurements of
background  normalization  (e.g.  Marshall  et  al.   \cite{Marshall};
McCammon \& Sanders \cite{McCammon}; Garmire et al. \cite{Garmire}; Vecchi et al. \cite{Vecchi}).

In the soft X-ray regime ($\leq$1keV), there is a substantial
Galactic contribution to the diffuse background originating
from hot plasma in the Local Hot Bubble, the Galactic Disk and
the Galactic Halo. The integrated spectra of these components vary over
the sky due to large scale spatial structure and varying amounts of 
absorption. In addition there may be a more uniform component arising from an
extragalactic hot intergalactic gas phase  (e.g. Cen \& Ostriker \cite{Cen}).

Here we describe a set of template background data sets compiled from
high galactic  latitude pointings of  the EPIC cameras  on XMM-Newton.
We discuss the main features of the observed background and the problems
the background poses for extended  source analysis. Finally we compare
our results with previous measurements of the spectrum
and normalisation of the diffuse cosmic X-ray background.

\section{EPIC}
The three co-aligned mirror modules of XMM-Newton each have an imaging
camera  at  their  focus,  provided  by the  EPIC  consortium.   These
comprise  two different  technologies: a  conventional  CMOS CCD-based
imager enhanced  for X-ray sensitivity (Turner  et al.  \cite{Turner})
and a pn junction technology  multilinear CCD camera (Str\"uder et al.
\cite{Struder}).  The first  type  is located  behind  the two  mirror
modules  containing  Reflection Grating  Arrays.  Each  comprise of  7
individual CCDs, closely butted, with  a pixel size of $\sim$1 arcsec.
The PN  camera is located  behind the third, unobscured  telescope. It
comprises an  array of  12 CCDs  in a monolithic  silicon array  - its
pixels subtend about 4 arcseconds square.

The cameras offer angular resolution determined by the telescopes Full
Width  at  Half  Max (FWHM)  of  5  arcsec,  a  field of  view  nearly
30\arcmin~ in diameter, energy resolution of typically 100~eV (FWHM) and
an energy range $\sim$0.2--10~keV. Each  camera is provided with a set
of  optical  blocking  filters  to reject  possible  contamination  by
visible  light photons.  On-board  electronics select  events above  a
certain threshold  and transmit data in  the form of a  serial list of
event locations and energies.
  
\section{Data Selection}
\subsection{Field Locations}
For previous missions (e.g.  ROSAT, Plucinsky et al.  \cite{Plucinsky}
and ASCA, Gendreau et al.   \cite{Gendreau}) a measure of the internal
background  components was  attempted  with data  collected while  the
instruments   were   pointed   at   the   dark   hemisphere   of   the
Earth. Subsequent analyses of representative sky backgrounds were also
made by compiling data from nominally source-free fields.

The XMM-Newton pointing restrictions, determined by the solar array or
attitude  measurement sub-systems, prevent  accumulation of  data from
the  dark  Earth or  Moon.   Initial  concerns about  unrepresentative
camera shielding  configuration led us to believe  that data collected
with  a  closed instrument  door  would  not  be useful  for  internal
background characterisation, particularly  with respect to fluorescent
line emission. However, we  are currently compiling observations with a
closed  filter position,  and initial  results seem  promising (albeit
with  low  observational  efficiency  and hence  low  signal-to-noise  at
present).  In  the meantime we  resorted to compiling data  from blank
sky fields. To do so with  realistic signal-to-noise ratios for each of 4
major instrument modes would  impose an unacceptable penalty on usable
Guest Observer science programme time, therefore we concentrate on the
Full  Frame imaging  modes  which  are generally  used  for the  faint
extended objects and which are our  primary concern. We considered that to
minimise statistical uncertainties, the effective exposure duration in
our data sets should be an  order of magnitude longer than that of the
typical  Guest  Observer  exposure  ($\sim$30~ks). Again,  it  was  an
unrealistic proposition  to observe with  the 3 EPIC  optical blocking
filters for  $\sim$300~ks each,  specifically to obtain  these
data. Therefore we decided to make  use of a variety of Guaranteed and
Calibration Time observations of  ``blank'' fields to compile our data
serendipitously.  Co-addition  of the  multiple  fields  allows us  to
minimise   any   effects  of   ``cosmic   variance''  resulting   from
pathological field  sources or variations  in the local  diffuse X-ray
emission.  In  addition,  any   time  variability  in  the  instrument
behaviour will be diluted.

It  was found  that suitable  fields were  almost all  taken  with the
filter in the THIN position,  as this choice gives the maximum
detection efficiency when there  are no bright optical objects in
the field.  Whilst imposing a requirement for the THIN filter enhances the
ability  to obtain  a homogeneous
data set, it does impose  additional complications for those users who
have chosen to use a  thicker optical  blocking filter  in  their own
observations yet  still  need a reliable set of background
templates  (see  Sect.~\ref{sec:caveat}).   Nevertheless as  the  soft
diffuse X-ray component is more spatially and spectrally variable than
the harder X-ray emission components this was not the major driver for
our analysis.

Table 1 summarises the location of the fields selected.
\begin{center}
\begin{table*}

\begin{tabular}{l r c c c l r} \hline
RA &Dec &Date of &Duration & \nh &L$_{\rm II}$& B$_{\rm II}$\\
(2000)   & (2000)    &observation&(ks) &(10$^{20}$ atoms cm$^{-2}$)& & \\
\hline
02:18:00&-05:00:00&2000 July 31&60&2.5&169.7&-59.8\\
02:19:36&-05:00:00&2000 August 4&60&2.55&170.35&-59.5\\
02:25:20&-05:10:00&2001 July 3&25&2.7&172.3&-58.6\\
02:28:00&-05:10:00&2001 July 6&25&2.7&173.5&-58.2\\
10:52:44&+57:28:59&2000 April 29&70&0.56&149.3&53.1\\
12:36:57&+62:13:30&2001 June 1&90&1.5&125.9&54.8\\
13:34:37&+37:54:44&2001 June 23&80&0.83&85.6&75.9\\
22:15:31&-17:44:05&2000 November 18&55&2.3&39.3&-52.9\\ \hline
\end{tabular}
\caption{Summary of target locations compiled. The co-ordinates refer to
the nominal centre of the field of view.}

\end{table*}
\end{center}

\subsection{Data Generation}
The  data  sets  were  processed  using  the  pipeline  processing  of
XMM-Newton  Science  Analysis  Subsystem  5.2 (SAS), in  order  to  generate
calibrated event  lists for  each EPIC camera.  Part of  this standard
processing  removes the  effects  of bad  pixels  detected within  the
observation.  Although many  of these  are  fixed in  location on  the
detector (and  hence may be present  in a Guest  Observer's data set),
occasionally a  pixel will become temporarily bad  (repeated low level
signals  unrelated  to  X-ray  events).  These were  removed  from  an
individual observation where the repetition was typically greater than
0.5--1\% of CCD readouts. (see Sect.~\ref{sec:caveat})

The signatures of most cosmic ray events are removed within the instrument
before transmission of data to ground. To account properly for the ``dead time''
effect of detector areas affected by cosmic rays, and therefore not available 
for detection of valid X-rays, a correction factor is calculated within
SAS,  based on a housekeeping parameter which counts the number of 
over-threshold pixels per frame, which is then normalised according to 
previous analysis of the average size of cosmic ray events 
observed in the directly transmitted images collected early in the mission.

\begin{figure}
\begin{center}
\resizebox{\hsize}{!}{\includegraphics{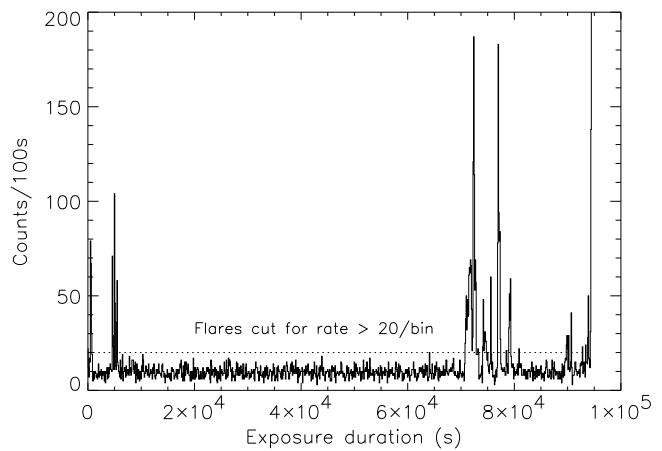}}
\caption{The light curve of the high energy ($\approxgt$~10~keV)
particle events detected in the MOS camera during a typical
observation. Intervals with count rates above 20 per 100~s bin were
rejected.}
\label{fig:rates}
\end{center}
\end{figure}

An important  background component which was  not properly anticipated
before  launch, is  a flux  of low  energy particles  (believed  to be
protons)  which  can  be  focused  by  the  mirror  systems  onto  the
detectors.  Within  the   magnetosphere  the  spacecraft  occasionally
encounters  concentrated ``clouds''  of accelerated  particles.  Their
signature in the detectors is of  bursts of events with few keV energy
deposition that  are stopped within  one CCD pixel. The  intensity can
reach more  than 100  times the quiescent  background rate  during the
worst  of  ``proton  flares'',  and these  intervals  are  essentially
unusable  for analysis  of  faint extended  objects.  We removed  such
occurrences in  our dataset by  using a screening procedure based on
the measured count rate of high energy single  pixel  events 
($\geq$10~keV).  Histograms of  such events collected  in time  bins of
$\sim$100s were  made and  data intervals with  count rates  $\geq$ 45
(20) events/bin in PN (MOS)  are rejected (see Fig.~\ref{fig:rates}).

Concerns that in this high altitude orbit, a low level of protons might
continually be illuminating the mirrors, have been  allayed with the information
gained from Chandra observations of the Dark Moon. In this case the recovered
spectra are statistically identical with a large selection of observations
made with ACIS outside the Chandra focal point (Biller et al \cite{Biller}). 

Next, we removed the signature of bright field sources. For each cleaned
exposure an image was formed in the 0.5--2~keV band (which is devoid of most 
image artefacts). The SAS task $EBOXDETECT$ was used to perform a simple 
sliding-box point source detection, without an exposure map, with a box size of
12\arcsec~ and a task detection threshold of $\sim$30--40. Depending on the
content of each field, approximately 10 objects per field were identified
and a circular exclusion region of 25\arcsec~ radius was applied
at each source position.   The resulting source-exclusion threshold cannot be
directly associated with a source flux level, but nevertheless analysis of
the individual fields confirmed the estimate,  based on published
LogN-LogS curves (Hasinger et al.   \cite{Hasinger}), that the cores of 
sources of flux
brighter than about 1 - 2 $\ $ 10$^{-14}$ erg cm$^{-2}$ s$^{-1}$ (0.5--2
keV) have  been excised. According to the latest telescope PSF
calibrations the above process fails to remove about 10 -- 20\% of their flux 
(almost energy independent), implying about 4 10$^{-13}$ erg cm$^{-2}$s$^{-1}$ deg$^{-2}$ of the bright point source flux (2 -- 10keV) remains.
This is addressed in Sect.~\ref{sec:xgal}.

It should be  noted that where there is no coincidence of point sources from
field to field, this excision would lead to a local depression of counts in
the final data set at  only the 10--15\% level. Conversely a suppression of
the remaining brightest sources by the co-added fields leads to a similar
level of dilution.  

Finally the screened data were co-added, and the exposure and GTI
({\it Good Time Interval}) extensions
of the FITS data files carefully added together so that an event list
suitable for use by the \xmm~SAS and other standard X-ray analysis tools
(e.g., XSPEC) was created. At the time of
writing, these data sets are available  from the XMM-Newton Science Operations
Centre ftp site $ftp://xmm.vilspa.esa.es/pub/ccf/constituents$.

We noted that, despite the above standard recipe for filtering proton flares,
the  background rate as measured in the 1--10~keV band exhibited some remnant
flares. This  implies that at the lower energies, the proton flares can turn
on or off at a different rate to the main flare component. We attribute this
to the existence of lower energy protons at the edges of the encountered
proton ``clouds''. We have not chosen
to force further stringent data cuts in the background template files, so as
to allow the general observer the option of applying additional selection on
the template files so as to match his/her own requirements.  However it was
found that in analysing the spectrum of diffuse X-rays in the field, the 
recovered spectral slopes steepened with more stringent flare mitigation.
Therefore before performing the spectral analysis described in
Sect.~\ref{sec:specan} we repeated the filtering procedure but with an
energy range of 1 -- 10~keV and a count rate threshold of
$\approxlt$1.15 s$^{-1}$ for the MOS cameras.

\section{Data Characteristics}
The useful exposure time collected is about 400~ks. With such high
signal-to-noise it becomes possible to identify some peculiar instrumental
features. These are worthy of note since it will be important to take
these into account in science analysis.

\subsection{Fluorescent Emission Lines}
\begin{figure}
\begin{center}
\resizebox{\hsize}{!}{\includegraphics{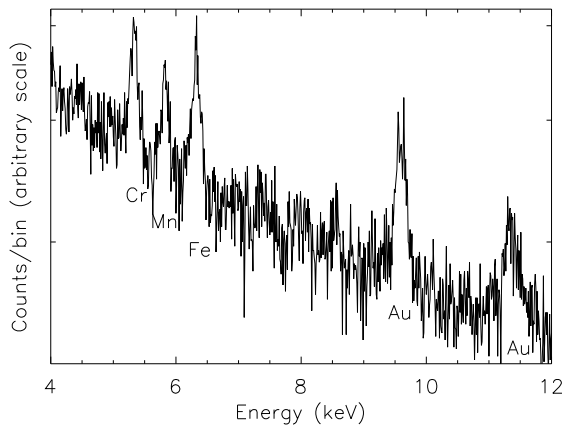}}
\caption{Background spectrum from the  MOS camera showing
fluorescent emission from the camera body materials.}
\label{fig:mosmetal}
\end{center}
\end{figure}

\begin{figure}
\begin{center}
\resizebox{\hsize}{!}{\includegraphics{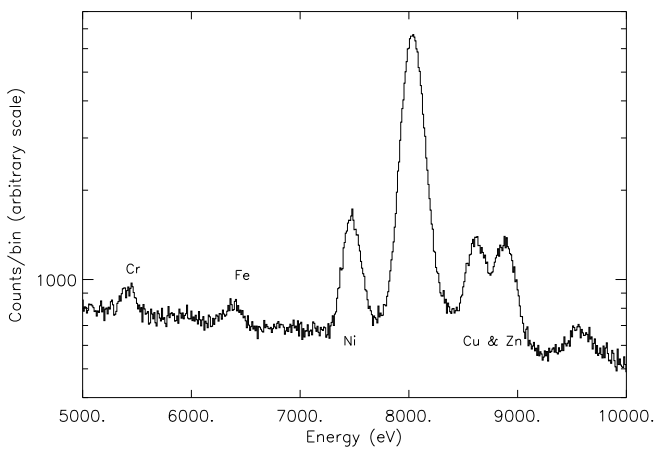}}
\caption{Spectrum from the  PN camera around the  region of the Cu K 8.05~keV emission line, showing fluorescent emission from other camera body materials. }
\label{fig:pnmetal}
\end{center}
\end{figure}

\begin{figure}
\begin{center}
\resizebox{\hsize}{!}{\includegraphics{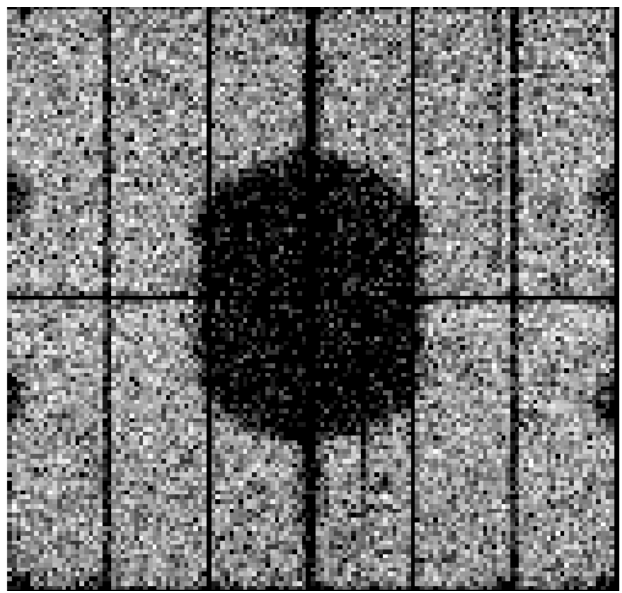}}
\caption{Image from the PN camera formed in the energy band of Cu K
fluorescent emission. Vertically and horizontally we see gaps where edge 
pixels have been removed. }
\label{fig:Cuimage}
\end{center}
\end{figure}

The passage of charged particles through the cameras is associated with
generation of fluorescent X-ray emission. This emission is most clearly seen
in the form of emission lines with energies characteristic of the camera body
materials
(Al and stainless steel components for example). The construction of
the two EPIC camera types is quite different, leading to substantially
different manifestations of these features.
\begin{figure}
\begin{center}
\resizebox{\hsize}{!}{\includegraphics{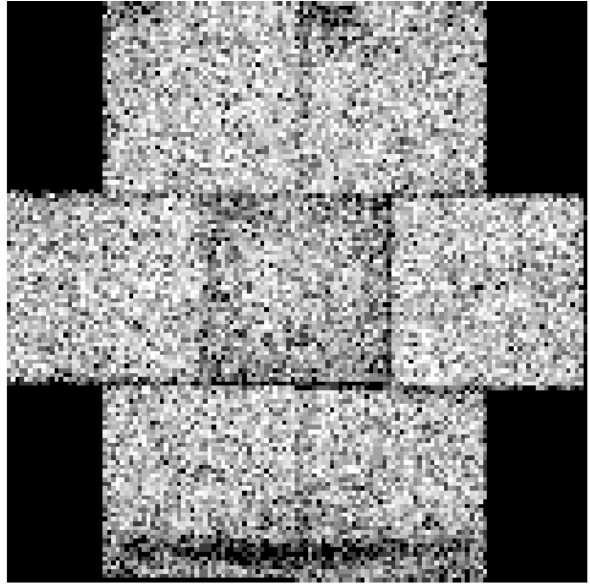}}
\caption{Image from the  MOS camera formed in the energy band of Al K
fluorescent emission. Gaps between the chips are barely visible but some 
shadowing features at the top and bottom are evident (due to cut outs in the
camera body for internal calibration sources). The rim areas of the 
central CCD are significantly dimmer than those of  surrounding chips, because
the CCD lies slightly below the others, and there is thus shadowing of any
emission from the camera body above the focal plane. }
\label{fig:Alimage}
\end{center}
\end{figure}
\begin{figure}
\begin{center}
\resizebox{\hsize}{!}{\includegraphics{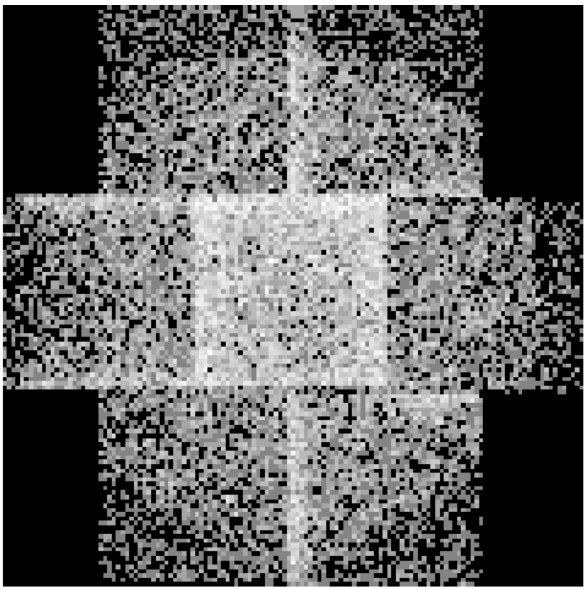}}
\caption{Image from the  MOS camera formed in the energy band of Si K
fluorescent emission. The image is brighter  along the inter-chip gaps due to 
the collection of Si K photons from the rear of chips which are located
above their neighbours.}
\label{fig:Siimage}
\end{center}
\end{figure}
Figs.~\ref{fig:Alimage} and ~\ref{fig:Siimage} show coarse binned 
images of a MOS camera in 
bands centred around the energies of Al K and Si K emission  respectively.
The outer 6 of 7 CCDs detect more Al K radiation due to their closer proximity
to the Al camera housing. Si K emission however is concentrated
along the edges of some CCDs. This is attributed to detection of Si K X-rays
escaping from the back substrate of a neighbouring CCD placed slightly
forward and overlapping the subject chip. (This physical stacking arrangement
of the CCDs was necessary to maximise their close packing to reduce dead space
between the arrays). Finally there are a number of much less intense emission 
line features at higher energies (Cr, Mn, Fe K and Au L lines for example,
see Fig.~\ref{fig:mosmetal})
generated by trace elements of the camera metal bodies.
In contrast the PN camera is monolithic and planar, so sees no Si K emission
which is self absorbed in the large pixel (150~$\mu$m) dimensions. In addition
to the Al K background there is a relatively intense contribution from 
energies around the Cu K line of 8.048~keV (Fig.~\ref{fig:pnmetal}). What is especially 
notable is that
this emission is spatially variable (Fig.~\ref{fig:Cuimage}). The central 
``hole'' mirrors
rather precisely the construction of the printed circuit board carrier on
which the CCD array is mounted.


The consequences of ignoring the spatial variation of these background
features could be dramatic. XMM-Newton is possibly the observatory
of choice for spectrally-resolved imaging of large clusters, for example to
map radial temperature and element distributions. However, the variable Al
and Si background lines could compromise abundance determinations of 
cluster emission lines with moderate redshifts, while the variable high energy
background would bias temperature measurements at large radii. These 
difficulties 
should be alleviated if the proposed background templates prove to be 
representative.
    
\subsection{Unrejected Particle Background}

The charged particle induced events which are not rejected by the on-board
or  ground  processing give rise to a background component that is
relatively  constant in  spectrum and which shows little variation
across the detector. 

 A detailed study of the XMM-Newton background environment has been carried 
out (Dyer et al \cite{Dyer}). Using the CREME software (Adams et al \cite{Adams}), the XMM-Newton operational orbit (60,000 -- 100,000 km) was
predicted to experience close to interplanetary galactic cosmic ray spectra
at solar minimum conditions. A raw rate of 4.4 cm$^{-2}$ s$^{-1}$ would 
reduce by a factor $\sim$2 at solar maximum. Geomagnetic shielding has a small 
attenuation effect on the protons in the XMM-Newton orbit, below a few hundred MeV, but this affects the {\em total} fluence at less than 10\%.
The  measured rates in all  the 3 different sorts of CCD cameras on  \xmm 
(including the CCDs in the RGS instrument)  
since launch are in the range 2--2.5 cm$^{-2}$ s$^{-1}$, which is entirely 
consistent with the expected solar modulation of the cosmic ray rate.

The  mean charge  deposition in  both  EPIC cameras  due to  minimally
ionising particles  should be $\gg$10~keV  per particle and,  with
such events typically crossing multiple pixels, the signature of the charged
particles   should   be  easy   to   distinguish   from  valid   X-ray
events. Actually,  due to the Landau energy  distribution of deposited
energy the  rejection efficiency is  expected to be $\sim$99\%  in the
MOS cameras (Lumb  \& Holland \cite{Lumb}). There is  also a component
expected from  the secondary $\gamma$-rays and e$^{-}$  excited by the
particle interactions in the surrounding spacecraft.

\begin{figure}
\begin{center}
\resizebox{\hsize}{!}{\includegraphics[angle=270]{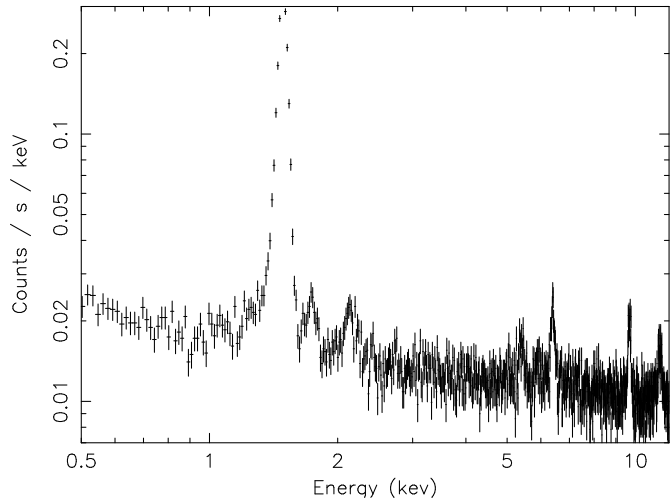}}
\caption{Spectrum from the  MOS camera in the region outside the field of
view. The flat spectrum is consistent with 99\% rejection of cosmic ray
events. }
\label{fig:cosmicray}
\end{center}
\end{figure}

In the case of the MOS devices, there is a substantial active CCD area
outside  the nominal  field of  view defined  by the  optical blocking
filter. In this region one  can assume that there are no contributions
from  sky photons or  soft protons  focused by  the mirrors,  and this
should      represent      the      true     internal      background.
Fig.~\ref{fig:cosmicray} shows this  measured internal spectrum in the
MOS1 camera below  10~keV (outside the field of  view). In this regime
the remnant count rate after selection for X-ray event characteristics
(the  SAS attributes  {\em  (\#XMMEA\_EM\&\&PATTERN in  [0:12])} )  is
0.026   events   cm$^{-2}$s$^{-1}$.  Of   this   0.021  $\pm$   0.0022
cm$^{-2}$s$^{-1}$ is in the  flat spectrum component with the remainder in
emission  line  components and a noise  component which increases to  lower
energies. The flat spectrum  count rate implies a rejection efficiency
of $\sim$99\%, as expected.

For the PN camera the flat spectrum component is about twice the intensity
of the MOS cameras. We attribute this to signals from the end of cosmic ray
tracks that are outside the spatial rejection mask, and deposit relatively
small energy signature in single pixels.
\subsection{Low-Energy Artefacts}
The most obvious features in the lowest energy band are effects of ``bad 
pixels''. In their simplest manifestation these represent a  pixel location
where a small  amount of leakage current is integrated during the CCD frame
accumulation time; this small signal is indistinguishable from the signal
generated by a valid X-ray event. A pixel which is consistently bad is
flagged for removal  on-board by loading a table of positions to be blanked
out. Occasionally a pixel can enter the transmitted data stream before it is
identified as bad. More  likely,
some pixels ``flicker'' on and off (Hopkinson  \cite{Hopkinson}) with low 
recurrence rate ($\leq$1\%), and  the efficiency for finding them post-facto
in the SAS pipeline is dependent on  many
factors, so that some such events may occur in the background template and
not the observer's data set and vice versa. 

There are additional features peculiar to each camera type. In the MOS
there  is  a  form  of  electronic interference  noise,  which  causes
patterns  of  spurious  low   energy  events  which  repeat  every  64
columns. Fortunately their signature  is easily recognised and removed
by  the  SAS pipeline  in  almost  all  observations. Also  there  are
occasions  when  a pixel  will  ``light  up''  for several  successive
frames. This  effect is attributed to  the trapping of  a large signal
from a cosmic  ray event, which subsequently is  released to the pixel
site over many seconds.

In  the  PN camera  there  are  occasional  blocks of  bright  pixels,
typically 4  pixels in  extent, along the readout direction. Their presence  varies from observation
to observation.   They arise from an  artefact of the  CCD offset bias
level calculation at  the start of each observation:  To calculate the
precise local  zero signal level the  average value for  each pixel is
determined from  typically 100  readouts. Although extreme  high value
samples  are excluded  from the  calculation, a  local excess  (due to
cosmic   ray   detection   for   example)  may   occur   during   more
readouts. Highly ionising events may also cause an electronic baseline
droop, creating a local decrement in the calculated tables. The effect
of  one discrepant  sample of  value  several keV,  averaged over  100
samples is  still enough  to cause a  noticeable shift in  the locally
applied detection  threshold.  The  actual readout and  calculation is
performed in sets of 4 CCD rows at a time, and thus repeated blocks of
typically  3-by-4 pixels  might be  expected to  exceed  the threshold
during  the subsequent  science  exposure readouts,  with much  higher
probability than normal.

For energies $\leq$200 eV there is evidence of streaks at the edge and
near the middle  of the array. Some of these can  be attributed to the
same occurrences of electronic  baseline droop, and/or charge transfer
loss  in  readout  that  cause  a false  event  signature  (especially
occurring in the first few pixels of the next frame readout).

\section{Spectral Analysis}  \label{sec:specan}
\subsection{Internal Background} \label{sec:internal}
The  high signal-to-noise  data set  described allows  the  diffuse X-ray
background  parameters to be  compared with  the estimates  made using
other observatories.  In this work  we restrict ourselves  to analysis
with the  MOS cameras only, because  the spatio-spectral complications
of the internal background can be reasonably well corrected using data
from  outside the  focal plane  area.  Similar subtraction  in the  PN
camera will be  the subject of further work,  when sufficient internal
background  data  can be  collected  during  the on-going  calibration
programme.  Furthermore, the MOS  CCDs have nearly identical responses
across the focal plane, and  the average response can be determined by
application of  a simple mirror vignetting  correction. Conversely for
the  PN, the  response varies  along each  CCD due  to the  effects of
charge transfer inefficiency, and the determination of a linear change
in response with position and  a radial vignetting function requires a
rather   complicated   weighting  function   which   is  still   under
development.

\begin{figure*}
\begin{center}
\resizebox{\hsize}{!}{\includegraphics[angle=270,scale=0.5]{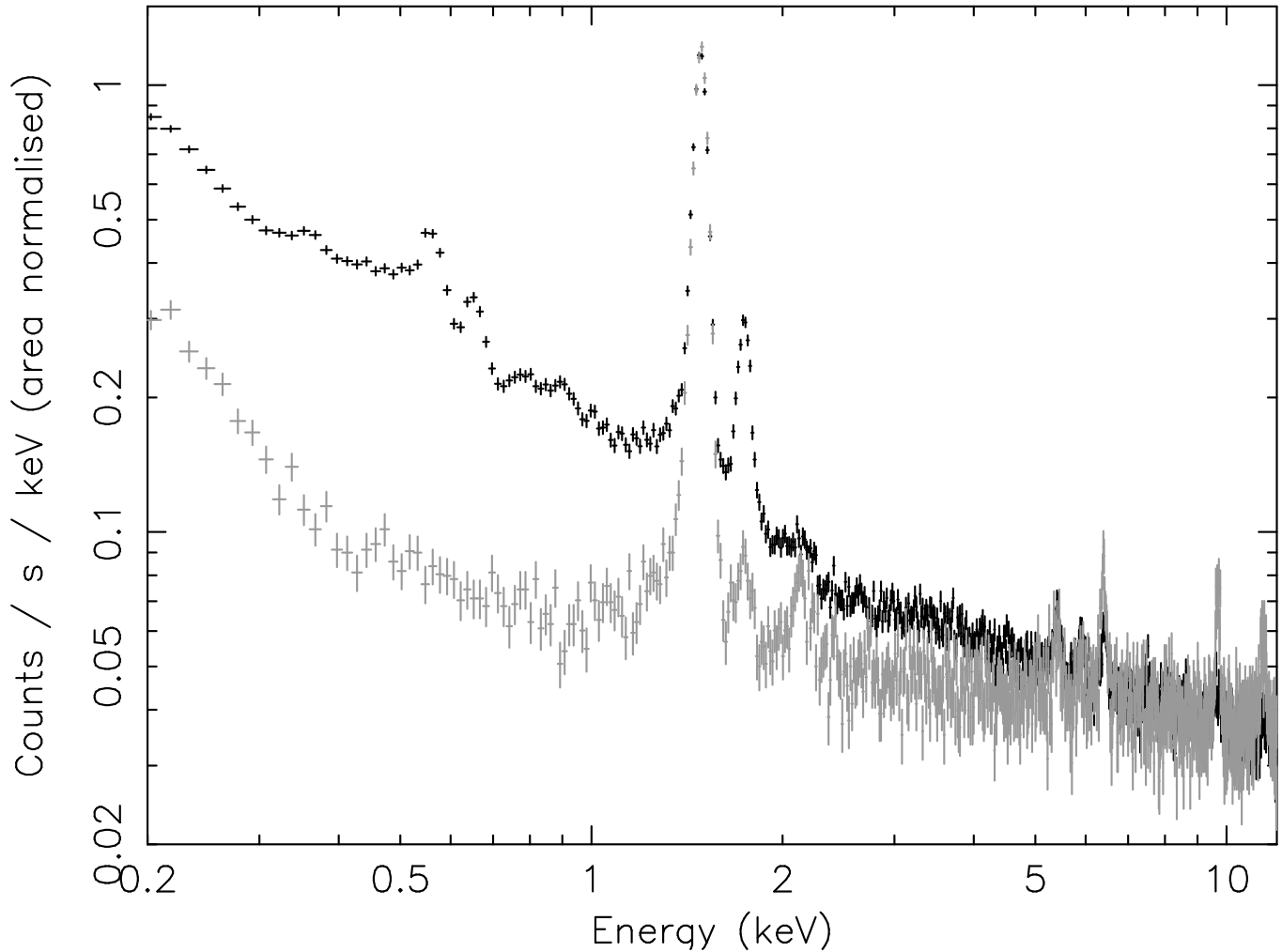}}
\caption{Comparison of the internal background (grey) and total background
spectrum (black) in MOS 1 camera. }
\label{fig:cxbint}
\end{center}
\end{figure*}

However,  even in  the  case of  the  MOS, analysis  using the  normal
technique of an area-scaled background subtraction was complicated. It
was  found that  e.g.,  the rather  variable  Al \&  Si emission  line
intensities have  an associated component of  redistribution of photon
signal to  lower energies.  The  intensity outside the FOV  is clearly
different than inside the  field, and therefore, even ``ignoring'' the
Al  K  line region  from  fitting  process,  after subtraction,  would
significantly  distort  the whole  sub-keV  spectral region.   Similar
arguments,  but  with less  dramatic  magnitude,  apply  to the  other
internal  fluorescent  lines.  In  Fig.~\ref{fig:cxbint} we  show  the
comparison of the internal background component from {\em outside} the
field of view (but scaled with appropriate area  correction) with the
total spectrum  (including the CXB)  from the central field of
the detector (specifically a central $\sim$13  arcminutes radius region).
The  internal component  dominating  above 5 keV
seems to  be scaled correctly via  the ratio of  in- and out-of-field
detector areas,  but the emission  line intensities at  Al K and  Si K
require a  different  scaling  (this  is partly masked by  the
diffuse cosmic X-ray background  component) consistent with
spatial variations noted earlier.  Therefore we  decided to model
the   internal  background,  with   multiple  Gaussian   functions  to
characterise these emission lines,  superimposed upon a broken power-law to
describe the continuum due to unrejected particle backgrounds, and low
energy noise.  The hard component of unrejected  cosmic rays manifests
as  a power law  of counts versus energy  of $\sim$E$^{-0.2}$,  turning up  at about
1~keV to a power law $\sim$E$^{-0.8}$ (this is very similar in form to models
used to describe internal background in the ASCA CCD cameras).  In the 
spectral fitting
described  below, the  fluorescent emission  line energies  and widths
were   fixed  to   the  values   determined  in   the  out-of-field
component but the normalisations of the lines were allowed to vary to
compensate for spatial  variations. The normalisation of the
broken power-law  component was fixed according to the ratio of
the collection areas.

Yet a  further complication  is that the  charged particle  events are
collected  unevenly across the  array. After  collection in  the image
area the data  are shifted rapidly into a  shielded store section, and
read-out relatively  slowly ($\sim$2.6 s)  so that rows farthest  from the
readout node spend  $\sim$50\% longer in the array.  The pixel size in
the store section  is less than 1/3 of the physical  size of pixels in
the  image section,  so the  actual range  in  equivalent accumulation
times is $\sim$15\%.  In addition the smaller pixel  size allows for a
much enhanced rejection ratio  to particle events collected during the
store  readout  than  those  collected  in the  image  section.  As  a
consequence of  the orthogonal readout  organization of the  MOS CCDs,
only  2 of the  6 outer  CCDs have  a potential  spatial bias  in this
cosmic  ray collection  efficiency, hence  the  expected normalization
error in the unrejected cosmic ray component is $\ll$5\%. We confirmed
this spatial  invariance by examining  a selection of data  sets taken
with the  CLOSED filter  position, and indeed  find no evidence  for a
large  spatial  variation  in  the  background  {\em  except}  in  the
fluorescent emission lines.

At energies $\leq$5~keV the CXB  dominates, with a clear signature of
low energy emission lines at E$\leq$1 keV, which are not present in the
internal background but most likely originate in  Galactic thermal
emission.

\subsection{Cosmic X-ray Background}

Using the spectral model of the internal background described in the 
previous section, we have proceeded to fit the {\em full background 
spectrum} measured in the EPIC MOS cameras by including additional spectral 
components representative of the diffuse CXB. Specifically
we  adopt   an  empirical CXB model consisting of two optically-thin
thermal components (MEKAL, Mewe   et  al. \cite{Mewe}) plus a power-law 
component. The former represent the soft CXB produced by hot plasma
located in the Galactic disk and halo whereas the latter models the hard 
CXB of extragalactic origin which, most likely, is due to the integrated 
emission of faint unresolved AGN.  

Other details of the spectral analysis are as follows. Since the 
metallicity  of the  thermal component could not be well  constrained,  
the plasma abundances  were  fixed at  solar  (Anders  \& Grevesse 
\cite{Anders}).  Also the exposure time  weighted hydrogen column 
density  (Dickey \& Lockman \cite{Dickey}) averaged over  all our fields
was  calculated  to be  1.7  10$^{20}$  cm$^{-2}$  and fixed  at  this
value. We have considered an  overall estimate of 2\% systematic error
is  appropriate  to  characterize  the current  level  of  calibration
accuracy, except for  the region around the O  absorption edge feature
in the CCD  detection efficiency, to which we  assign a 5\% systematic
uncertainty (450 eV--600 eV).  All spectral bins below an energy of  
200 eV were excluded.

In the spectral fitting we use response  functions which account for the  
average effective area, weighted according to off-axis angle.
 The vignetting calibration of the MOS cameras is complicated by 
an understanding of the azimuthal dependence of the  blocking fraction  of 
the  Reflection Grating  Spectrometer array (den Herder et al.  
\cite{denHerder}) modules in their telescope beams. 
Nevertheless as  we have azimuthally  {\em averaged}
this  factor in  these observations,  we believe  that  the uncertainty
of the vignetting
function  measurement ($\sim$1.5\% for E$\leq$~8keV at 10.5 arcminute field
angle -  Lumb et  al.
\cite{LUMBvig}) represents an upper limit to the uncertainty on the average 
effective area weighting applied in the analysis.  

Fig.~\ref{fig:spectrum}  shows   the  measured CXB spectrum (after 
subtracting the best-estimated of the internal background) in the 
0.2--10 keV band, our derived  best-fitting CXB spectral model and
the fitting residuals. The corresponding parameter values
plus errors are summarised in Table 2. The reduced $\chi^{2}$ of
the fit was \rchisq = 1.17  (566 degrees  of freedom).

The background normalisations quoted in Table 2 ignore the contribution of
out-of-field scattered X-rays. X-rays at angles  $\sim$1 degree off axis 
have  a small probability of
reaching the focal plane after  only one reflection (off the hyperbola
mirror)  rather   then  a  true  focus  from   the  double  reflection
geometry. For  example, such effects were  a considerable complication
in the  ASCA observatory since they  gave rise to roughly  30\% of the
signal. In  XMM-Newton, considerable effort  was put into  designing a
baffle  which minimizes  this effect.  Ground measurements  at optical
wavelengths have confirmed  its efficiency.  At the time
of  writing the  detailed  in-orbit X-ray  calibration is  incomplete.
However measurements  of the Crab Nebula observed  off-axis, show that
the  X-ray leak  is  within a  factor of  2  of predicted  (1.5 
10$^{-3}$). With a rather small energy dependence ($\sim \pm$20\% from
1 -- 10~keV) of the X-ray leak, this 
leads us to estimate that the contribution of diffuse flux
gathered from out-of-field angles  of 0.4 --  1.4 degrees  is of  order 
7\%  of the  true focused in-field signal, and  the associated systematic 
error (due largely to the energy dependence) is $\pm$2\%.

\subsection{Extragalactic Cosmic X-ray Background}\label{sec:xgal}

The hard power-law component representing the extragalactic CXB
has a measured photon index of $\Gamma$=1.42$\pm$0.03 (90\% 
confidence limits for one parameter) with a normalisation at 1 keV of
$8.44 \pm 0.24 \rm~photon~cm^{-2}~s^{-1}~keV^{-1}~sr^{-1}$ (corrected
for out-of-field scattering). The corresponding  2--10~keV flux per square 
degree is 1.64 10$^{-11}$ erg cm$^{-2}$ s$^{-1}$ deg$^{-2}$.

\begin{figure*}
\begin{center}
\resizebox{\hsize}{!}{\includegraphics[angle=270,scale=0.5]{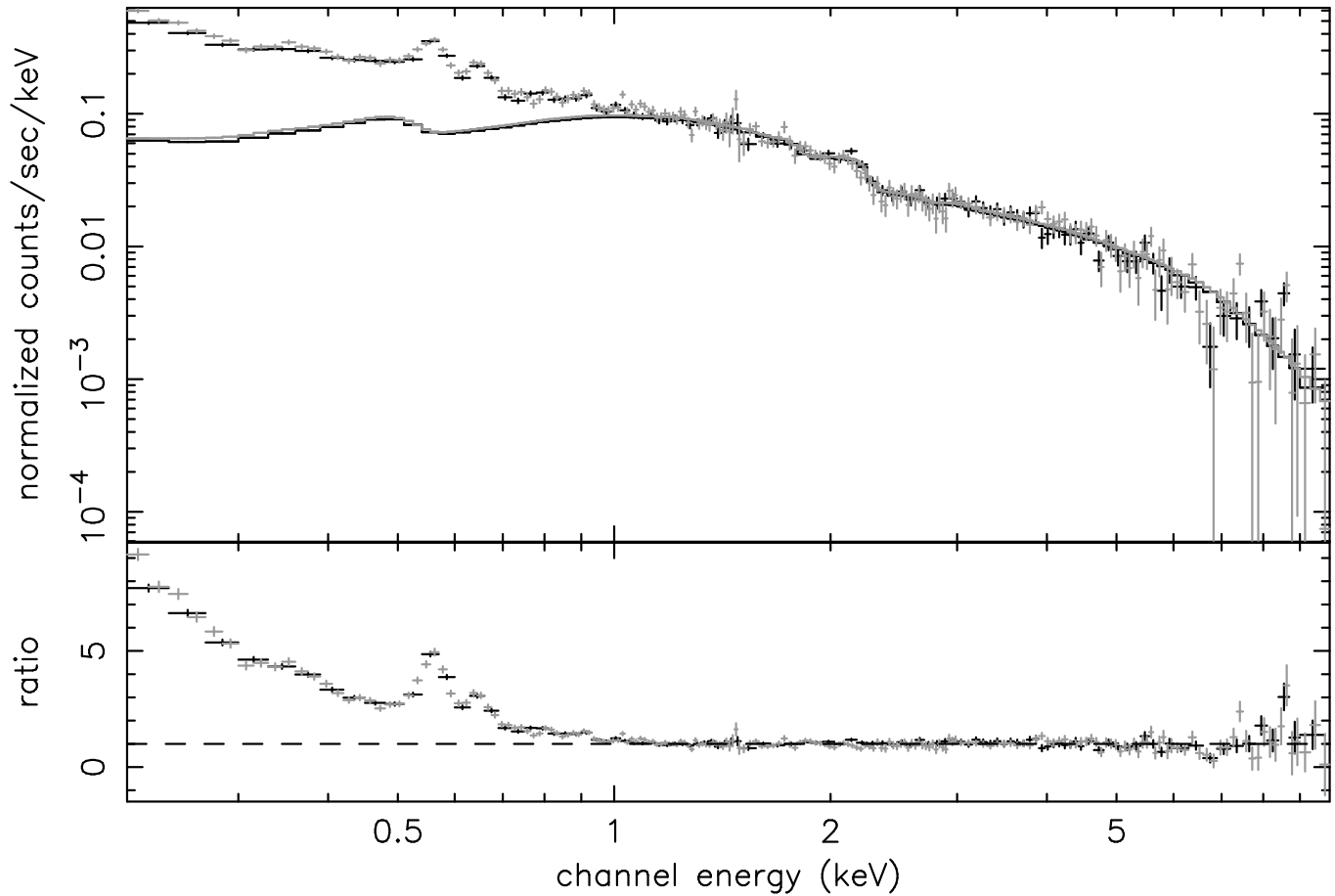}}
\caption{Comparison  of the overall  CXB measured  in the  2 MOS cameras
with the power-law extragalactic  component. The lower panel shows the
ratio   between   measured   background   and   a   power   law   with
$\Gamma$=1.42.  At the  highest energies  there is  evidence  for poor
subtraction of fluorescent emission lines in Ni, Cu and Au. }
\label{fig:xgal}
\end{center}
\end{figure*}

Fig.~\ref{fig:xgal} shows the contribution of the power-law component to
the whole spectrum; the residuals illustrate very clearly that the
breakpoint between the hard extragalactic power-law  and the soft CXB
emission of Galactic origin occurs close to 1~keV. This is a spectral
range where many instrument technologies overlap. For example,
missions flown in the past utilising either thick-windowed collimated gas
proportional counters or X-ray imagers with CCDs were mainly sensitive above
$\sim 1$ keV.  Conversely, the telescope and thin-windowed proportional
counters  of ROSAT were tuned to the detection of X-rays below $\sim 2$ keV.
Nevertheless, considerable evidence has accumulated for a soft excess of
Galactic origin in the CXB spectrum (eg. Bunner et al. \cite{Bunner};
Garmire et al. \cite{Garmire}). The fact that the EPIC cameras identify the spectral
break in the CXB in such a convincingly and unambiguous fashion does,
however, demonstrate the importance of broad spectral coverage for CXB
measurements. Presumably limited bandpass/energy leverage has also been a
factor in producing discrepancies of up to 30\% in published estimates of
the normalisation of the extragalactic background.

In order to compare our background measurements with other all-sky
averages (e.g. Della Ceca et al.  \cite{Della}) we must account for
the bright source component that has been removed from our data
( {\it i.e. the bright resolved sources that we have specifically
excluded from our summed spectra}). Currently the most secure 2--10 keV source 
count data in the bright source regime comes from ASCA and BeppoSax.  We 
have used the log N - log S curves  of  both Ueda   et  al. 
(\cite{Ueda}) and  Cagnoni et al.  (\cite{Cagnoni})  for the 2--10~keV band. 
We convert our lower flux limit (for the excluded sources) from the 
0.5--2~keV band to 2--10~keV via an AGN spectrum appropriate to the flux 
range ($\Gamma \sim$1.7). Thus we integrate the counts from  
3 10$^{-14}$ to 10$^{-12}$  erg cm$^{-2}$s$^{-1}$ (2--10~keV). This  
leads to a bright source correction factor of  6.0 10$^{-12}$ erg 
cm$^{-2}$s$^{-1}$ deg$^{-2}$.  The estimated uncertainty in that value is
driven by the possible range of 2 in the lowest flux from which we integrate 
the source counts ($\sim$2 -- 4 10$^{-14}$  erg cm$^{-2}$s$^{-1}$ in 
2--10~keV), and is estimated to be 2.0 10$^{-12}$ erg cm$^{-2}$s$^{-1}$ deg$^{-2}$.

When we combine our unresolved hard CXB intensity with the estimated
contribution of bright sources we obtain a value for the total intensity
of the hard CXB of  2.15 $\pm$0.26 10$^{-11}$ erg cm$^{-2}$ s$^{-1}$ deg$^{-2}$.
 Table~\ref{tab:error} lists the estimated uncertainties in computing this value. 

\begin{table}
\begin{tabular}{l c c l}
Component	&Value&Error&\\ \hline
Normalisation & 9.03 & 0.24&ph keV$^{-1}$cm$^{-2}$s$^{-1}$sr$^{-1}$\\
Avg. Vignetting     &0.68&0.01&In response fn.\\ 
Stray Light & 1.07 &0.02&Energy dependence\\
Bright Sources & 6.0 & 2.0 & x10$^{-12}$  erg cm$^{-2}$s$^{-1}$deg$^{-2}$\\
\end{tabular}
\caption{Summary of the components, and associated error estimates, used
in correcting for the final extragalactic background normalisation}\label{tab:error}
\end{table}
 

An independent check on the derived hard CXB intensity was made using data
{\em only} from inside the field  of view.  This was  accomplished  via  
maximum-likelihood  fitting  (Crawford et al \cite{Crawford}) of  vignetted  and unvignetted 
components to the  measured spatial distribution of events over the 
field of view.  In  this  case  a fit  was  made  for  an
un-vignetted  (cosmic-ray background)  and  a  vignetted (true  X-ray
photons)  image component,  with  a vignetting  function  of the  form
described by  the \xmm~  calibration database. Images  were constructed
from  both MOS  datasets in  150 eV  wide bands,  and  fit individually;
uncertainties  were determined during  the fits  using the  $\Delta C$
statistic  (Cash  \cite{Cash}).  The  maximum likelihood  fitting  was
unable to constrain  the vignetted component above 5  keV.  The fitted
intensities of the vignetted  component were then averaged between the
two  MOS detectors,  and  cast as  a  spectrum suitable  for use  with
XSPEC. Energy bins  corresponding to the strong instrumental  Al K and
Si  K emission  lines (i.e.  between 1.3  and 1.9  keV) and  where the
Galactic  background is  significant (below  1 keV)  were  excluded. A
power-law fit to the  spectrum of the vignetted background resulted in
$\Gamma$ =  1.34$\pm$0.10 (90\% confidence), consistent  with the best
fit value of $\Gamma =1.42$ obtained earlier.
Fixing the  slope at $\Gamma=1.42$,  and fitting the  normalisation of
our  vignetted spectrum  results in  a value  of ${\rm  A}_{\Gamma}$ =
9.1$\pm$0.4  {\rm photon~keV$^{-1}$ cm$^{-2}$  s$^{-1}$  sr$^{-1}$ (90\%
confidence), in excellent  agreement with the value given  in Table 2. 

\begin{figure*}
\begin{center}
\resizebox{\hsize}{!}{\includegraphics[angle=270,scale=.5]{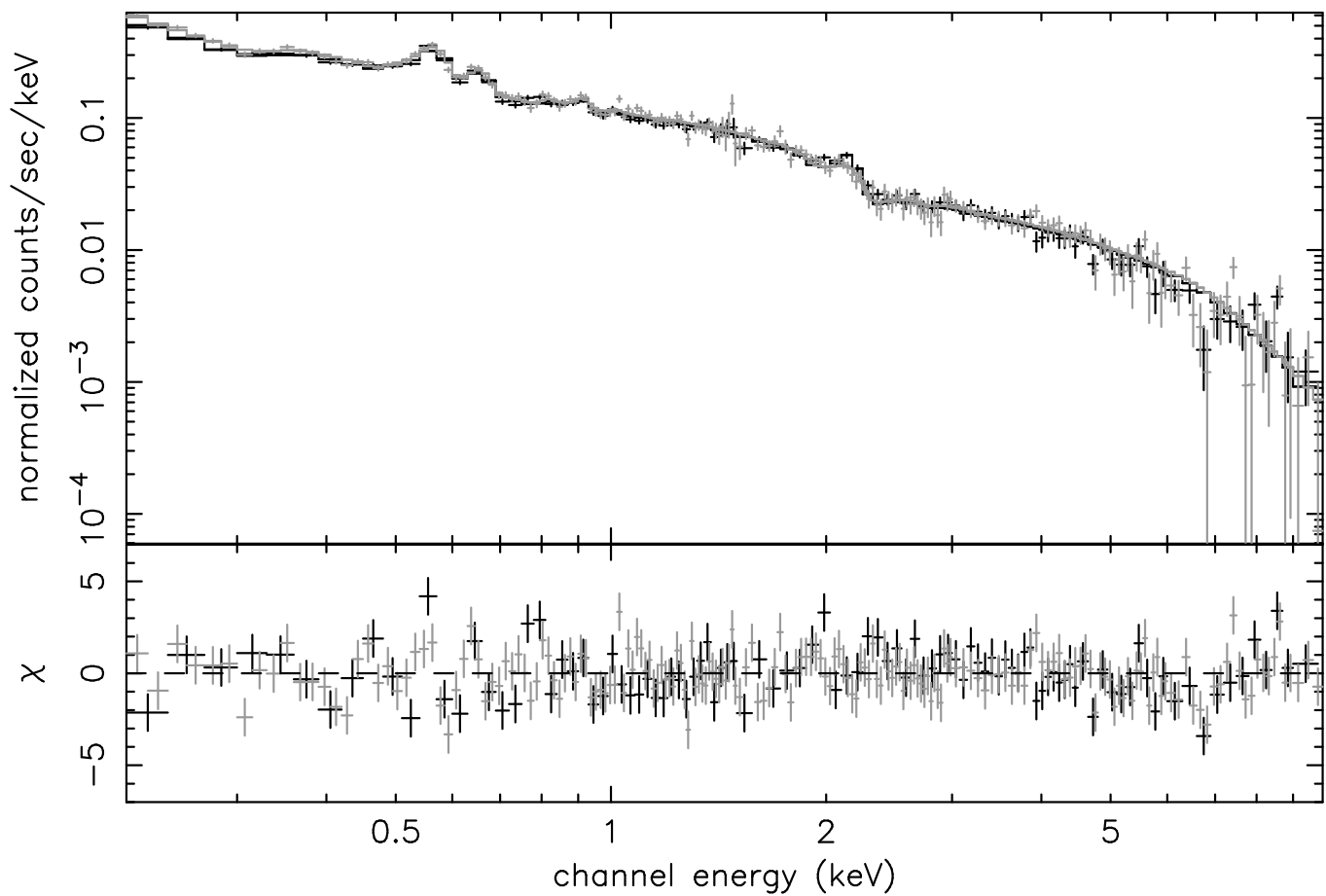}}
\caption{Best fit model and residuals to the overall cosmic diffuse X-ray
emission }
\label{fig:spectrum}
\end{center}
\end{figure*}
 To further validate our estimate of the corrections for missing bright
source flux, we made an independent fit on the whole EPIC MOS data set
with  bright  sources {\em included} and determined  a  2--10~keV intensity  
of 2.11 $\pm 0.11$  10$^{-11}$  erg cm$^{-2}$ s$^{-1}$  deg$^{-2}$. 
The correction for missed bright sources is estimated in this case to be 
only $\sim$1 10$^{-12}$ erg cm$^{-2}$ s$^{-1}$  deg$^{-2}$; thus this 
analysis gives a result  consistent with that noted earlier. The spectral 
slope in this case steepens to $\sim$1.45, as expected for the addition of 
the brighter steeper spectrum sources.

We also measured the brightest of the excluded sources, and determined
it  to  have  a flux  of  6  10$^{-13}$erg  cm$^{-2}$ s$^{-1}$  (2  --
10~keV).  This is  consistent with the source  count prediction  for a
total field of  1.3 square degrees, confirming that integration to 
higher fluxes from this point requires a minimal correction.

Our derived flux of the hard CXB signal compares well with  the
BeppoSAX (Vecchi et   al.   \cite{Vecchi})   and   Wisconsin  
(McCammon  \&   Sanders \cite{McCammon}) values,  but is somewhat higher 
than the HEAO-1 value  (Marshall et al. \cite{Marshall}, although see Garmire et al. \cite{Garmire}). Our result implies that the fraction of  the
background in the 2--10~keV band that has been resolved into discrete 
sources in the recent {\it Chandra} 1 Ms observations is   $\sim$70 -- 90\%.
This highlights the fact that estimates of the fraction of the CXB resolved 
into discrete sources in narrow beam surveys depend sensitively on the assumed 
normalisation of the CXB spectrum  and the corrections applied for the
contribution of bright sources (which must be deduced from wide area surveys) -
see for example, Baldi et al. (2001). The effects of field-to-field cosmic 
variance (which as yet are largely unquantified)  may also be important.

\begin{table*}
\begin{tabular}{l c c l}
Component	&Best Fit Value&Error&\\ \hline
$\Gamma$	&1.42	& 0.03&\\
A$_{\Gamma}$	&9.03	& 0.24&ph keV$^{-1}$ cm$^{-2}$ s$^{-1}$ sr$^{-1}$\\
kT$_{\rm 1}$	&	0.204 & 0.009& keV\\
A$_{\rm 1}$	        &7.59		&1.2 &\\
kT$_{\rm 2}$	&	0.074& 0.003& keV\\
A$_{\rm 2}$	&        116		&32&\\
\end{tabular}
\caption{Summary of the best fit spectral parameters derived for the cosmic 
diffuse background components. Normalisations are quoted per steradian, but do
not include corrections for stray light and missing bright sources.}
\end{table*}

\subsubsection{Galactic background}

We find that the soft CXB spectrum is well modelled by two thermal
components with temperatures of 0.07 and 0.20 keV respectively (see
Table 3). Kuntz  \& Snowden  (\cite{Kuntz}) describe  a  re-analysis of  
ROSAT survey  data, utilizing  different energy-band  intensity  ratios. For
high galactic  latitudes they estimated that after  correction for any
Local Hot Bubble component, a best fit two temperature model with kT =
0.099$^{+0.054}_{-0.037}$~keV  and  kT  = 0.24$^{+0.08}_{-0.03}$~keV
provided  a good  description  of the  measured ROSAT spectrum.  This
agrees  within  the errors  of  our own  estimate.  We  note that  the
emission line  ratios provide the key information  for the temperature
diagnostics, but the truncation  of EPIC response below 200 eV prevents
an  accurate   determination  of  the  absorption   columns,  and  the
0.25~keV emission  (knowledge of which is needed to disentangle  the Local
Hot Bubble component from more distant disk and halo emission).

We made an  estimate of the sub-keV emission in  some of the different
fields noting  differences in the measured temperature  and flux. This
variability in  emission and lack of  detailed temperature measurement
capability   hampers  a  more   accurate  determination   of  Galactic
background properties.   For example, the measured  mean deviation from
field to field  of the 2--10  keV intensity is about 3.5\%,  consistent with an
isotropic  extragalactic background. On the other  hand the mean
deviation of 0.2--1 keV intensity  is about 35\% from field to field. This
is highlighted  in Fig.~\ref{fig:galactic} where we show  the best two
temperature fits to several  fields, with their related deviations. It
is clear that:

\begin{itemize}
\item substantially different temperatures and absorption values apply to the
different fields,
\item that the residuals are not coherent from one field to the next
indicating that the instrument calibration is {\em not} the major impediment
to the low energy modelling,
\item a simple two temperature model is not an adequate description of the
sub-keV emission.
\end{itemize}

The  whole  subject of  determination  of  the  soft Galactic  diffuse
component is beyond the scope of the current work.

\begin{figure*}
\begin{center}
\resizebox{\hsize}{!}{\includegraphics[ angle=270,bb=114 0 554 700,clip]{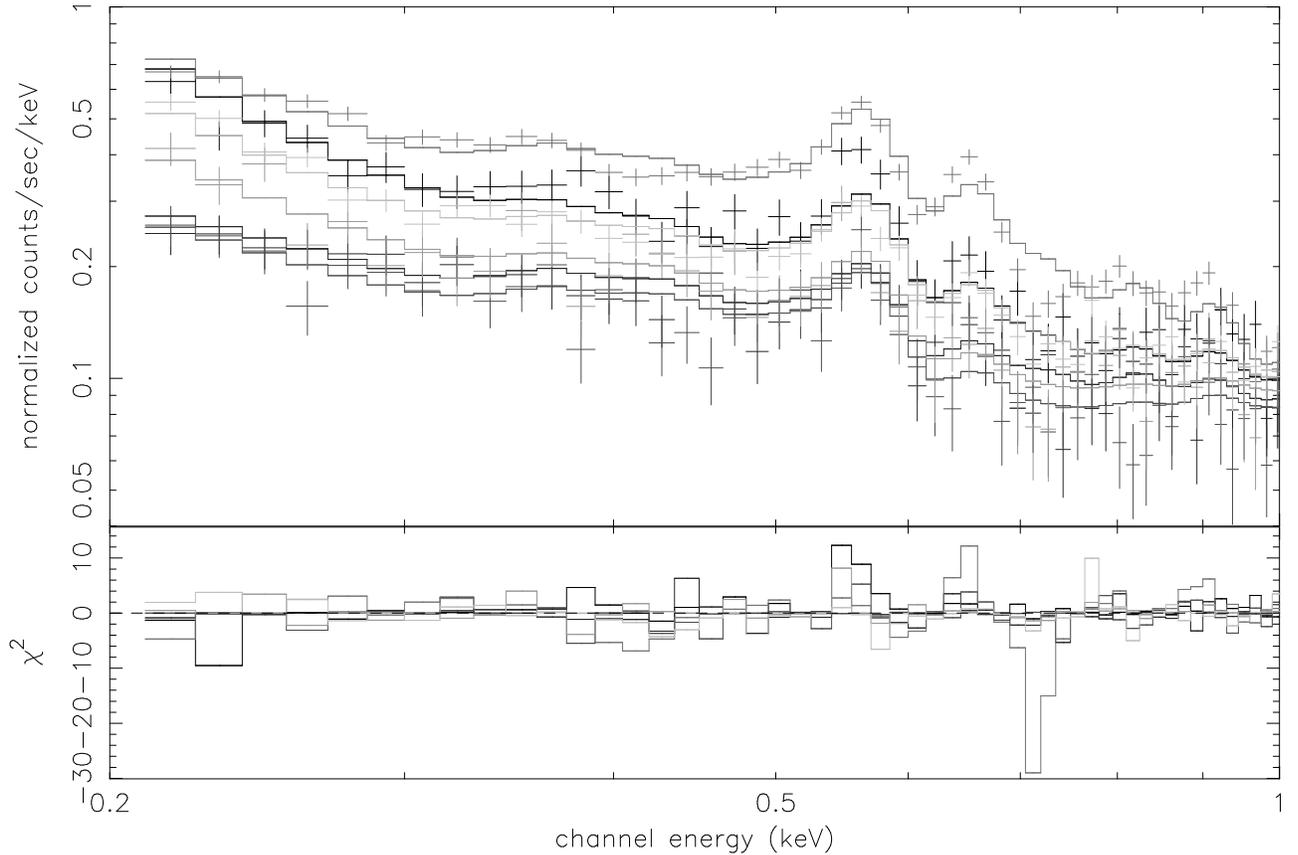}}
\caption{Comparison of the sub-keV spectra of a number of fields highlighting
the wide range of measured surface brightness.}
\label{fig:galactic}
\end{center}
\end{figure*}

\section{Caveats for Use} \label{sec:caveat}
The EPIC cameras allow a  choice of optical blocking filter to prevent
contamination by  optically bright  targets. For most  extended and/or
faint extragalactic targets, such contamination is negligible, and the
thinnest filter  can be employed,  as indeed is  the case for  all the
observations compiled  in this work.  However, if a thicker  filter was
used, the  transmission of diffuse background X-rays  (and any remnant
proton flux) will be reduced.  Therefore the observer should model the
differences  in soft  component  based on  the  response matrices  and
limited knowledge of the expected Galactic emission.

The observer must take care  to extract background from an appropriate
location  in the  FOV. Most  simply  the extraction  region should  be
defined in  detector co-ordinates to  match closely the region  of the
desired science  target. In some applications the  user would subtract
data in sky co-ordinates.  In this case the template event lists could
be recast  to mimic the  nominal pointing direction of  the observer's
field (for example using SAS task $ATTCALC$)

We note that despite the  standard recipe for filtering proton flares,
the  background  rate as  measured  in  1--10~keV  bands exhibits  some
remnant flares. This implies that at lower energies, the proton flares
turn on more slowly, yet before the main flare component. While a more
restrictive  proton  flare  screening  was applied  for  the  spectral
analysis of the background, we have not chosen to apply such stringent
data  cuts in  the  background  template files,  so  that the  general
observer can  apply an additional  selection of the template  files if
necessary.   We emphasise that  in analysing  the spectrum  of diffuse
X-rays in the  field, the recovered spectral slopes  steepen with more
stringent flare mitigation. Careful comparison of the recipes used for
GTI  creation must  be made.  Nevertheless,  it is  possible that  the
lowest level  proton fluxes are  spectrally variable, and  no complete
subtraction is possible.
 
Although  point sources  have been  removed or  significantly diluted,
careful  examination of  the  images derived  from  these event  lists
reveals  significant intensity fluctuations.  On scales  of arcminutes
appropriate  to  extended sources,  this  is  not  expected to  be  a
significant   problem,  and   indeed   representative  of  the  unresolved
background.  However,  if  the  user  extracts  spectra  from  regions
comparable with  the mirror point  spread function scale,  then manual
inspection is necessary to  guard against variations in counts arising
from treatment of point sources in the template files.

As noted previously there are particular defects to be expected in the
lowest  energy ($\leq$0.3~keV)  spectral ranges.  Furthermore,  at the
time  of writing,  the calibration  of  the EPIC  soft X-ray  spectral
response awaits  completion. The  transmission of filters  at energies
$\leq$250~eV  is difficult to  measure, the  CCDs' calibration  at the
ground synchrotron facility was not performed at energies $\leq$150~eV
and  the detailed redistribution  of signal  from photons  of energies
$\sim$1~keV into  partially collected events  in the softest  band was
also not  determined completely in  ground measurements. For  the time
being spectral analysis below 250 eV should be treated with caution.

For the highest quality determination of  the background appropriate to the
observer's own  data the  following steps should  be employed.   It is
intended to  provide tools within the \xmm~SAS  environment to achieve
this, but most steps can be performed manually (see for example \cite{Majero} ).

\begin{itemize}
\item Following  suitable flare screening, define  a background region
(B) and extract the observed spectrum (C$_{\rm back}$) from the observer's
data set. From  an identical region in the  template file the observed
spectrum (T$_{\rm back}$)  should provide a  measure of variability  of CR
component  by checking  count rates  for E$\approxgt$5~keV  and/or the
fluorescent emission line normalisations.
\item  To  estimate a  better  internal  background  spectrum for  the
observer's  data  set   (C$_{\rm inst}$),  determine  a  predicted  cosmic
background spectrum for  the observer's region based on  ROSAT All Sky
Survey maps,  hydrogen column etc.. An experimental  tool is available
at HEASARC web site to aid this (Sabol \cite{Sabol}).  Create response
matrices  for the  background  region  (here is  where  the effect  of
different  filters  can be  introduced).   Fold  this cosmic  spectrum
through  the response  matrices  to obtain  a  {\em predicted}  cosmic
component  for the  background  region, (C$_{\rm cos}$).  \\ C$_{\rm inst}$  =
C$_{\rm back}$ - C$_{\rm cos}$
\item A  similar approach with the  template files showed  that with a
weighted  average N$_{\rm H}$  of 1.7  10$^{20}\rm~cm^{-2}$, and  a 
0.2~keV thermal spectrum determined  from our  spectral fitting,  the 
HEASARC  tool  predicts a ROSAT
R$_{45}$ PSPC  count rate of 1.3~s$^{-1}$ in  a 144 arcmin$^{2}$ field,
and a  0.47-1.21 (ROSAT  band) flux of  1.67 10$^{-11}$  erg cm$^{-2}$
s$^{-1}$.  This could  likewise be  used to  make an  estimate  of the
internal background  of the  template file region  in order  to better
estimate  the  scaling factor  (K)  for  the  cosmic ray  component.\\
T$_{\rm inst}$      =      T$_{\rm back}$      -     T$_{\rm cos}$      and      K
$\sim$C$_{\rm inst}$/T$_{\rm inst}$\\
\item  Repeating the  same  exercise  for the  source  region in  both
template  and observed  data  sets  could lead  to  a {\em  predicted}
background data spectrum comprising the scaled internal component, and
the   predicted  galactic  component   with  the   appropriate  filter
responses.
\end{itemize}

\section{Conclusions}
Proper  treatment of the  background observed  in the  XMM-Newton EPIC
cameras is very  important for spectral and spatial  analysis of faint
extended objects.  The template  background files described  provide a
very   high   signal-to-noise  characterisation   of   the  high   energy
($\geq$1~keV)  background  suitable  for  such analysis.  The  sub-keV
background is consistent with models of the Galactic diffuse emission,
but is spatially and spectrally variable, and needs indirect arguments
to  provide  an  accurate  subtraction.  The  extragalactic  power-law
background component has been  measured, and is consistent in measured
power-law index  with previous data. The normalization  measured is in
good agreement  with, and provides an  important additional constraint
of  the fraction  of background  resolved into  point sources  by {\it
Chandra} and XMM-Newton.
\begin{acknowledgements}
All the EPIC instrument calibration team who have contributed to understanding
the instrument are warmly thanked for their efforts. During early developments
of the background template file activity there was helpful feedback from users
of the data for analysis of PV cluster targets. We wish to record the helpful
discussions with M Arnaud, U Briel, D Neumann, J Nevalainen, R Lieu and 
S Snowden. We thank the anonymous referee for useful comments.

\end{acknowledgements}


\begin{thebibliography}{}
\bibitem[1981]{Adams}
Adams, J.H., Silberberg, R. \& Tsao, C.H. NRL Memorandum Report, ``Cosmic ray effects on microelectronics'', Naval Research Lab, Washington, USA 1981.
\bibitem[1989]{Anders}
Andres, E. \& Grevesse, N., 1989 Geochimica et Cosmochimica Acta 53, 197
\bibitem[2001]{Baldi}
Baldi, A., Molendi S., Comastri, A. et al., 2001 submitted to ApJ astro-ph0108514
\bibitem[2002]{Biller}
Biller, B., Plucinsky, P. \& Edgar, R. Chandra X-ray Center Calibration memo, 01\_22\_02
\bibitem[1971]{Bunner}
Bunner, A.N., Coleman, P.L., Kraushaar, W.L., McCammon, D., 1971, ApJ, 167, L3
\bibitem[1998]{Cagnoni}
Cagnoni, I., della Ceca, R. \&  Maccacaro, T., 1998 Ap J 493, 54
\bibitem[1979]{Cash}
Cash W., 1979, ApJ, 228, 939
\bibitem[1999]{Cen}
Cen, R. \& Ostriker, J. P., 1999 ApJ 514, 1 
\bibitem[2002]{Cowie}
 Cowie, L. L., Garmire, G.P., and Bautz, M.W. et al, 2002 Ap J 566 L5 
\bibitem[1970]{Crawford}
Crawford D.F., Jauncey D.L., \& Murdoch H.S., 1970, ApJ, 162, 405
\bibitem[1999]{Della}
Della Ceca, R., Braito, V. and Cagnoni, I. et al., 1999, Ap J, 524, 674 
 \bibitem[2001]{denHerder}
den Herder, J. W., Brinkman, A. C., Kahn, S. M. et al., 2001, A\&A 365, L7
\bibitem[1990]{Dickey}
Dickey, J. M. \& Lockman, F .J. 1990, ARAA. 28, 215.
\bibitem[1995]{Dyer}
Dyer, C.S., Truscott, P.R., Evans, H.E. \& Peerless. C.L. Defence Research Agency Report, ``Analysis of XMM instrument background induced by the radiation environment in the XMM orbit'', DRA/CIS(CIS2)/CR95032, DRA Farnborough, U.K. 1995
\bibitem[1992]{Garmire}
Garmire, G. P., Nousek, J. A., Apparao, K. M. V. et al., 1992, Ap J 399, 694
\bibitem[1995]{Gendreau}
Gendreau, K., Mushotzky, R. \& Fabian, A., 1995, PASP 47, L5
\bibitem[2001]{Giacconi}
Giacconi, R., Rosati, P., Tozzi, P. et al., 2001, Ap J  551, 624
\bibitem[1998]{Hasinger98}
Hasinger, G., Burg, R., Giacconi, R. et	al	 1998
A\&A	329, 482 
\bibitem[1993]{HasingLS}
Hasinger, G., Burg, R.,  Giacconi, R. et al. 1993 A\&A 275, 1
\bibitem[2001]{Hasinger}
Hasinger, G., Altieri, B., Arnaud, M. et al., 2001, A\&A 365, L45
\bibitem[1992]{Hopkinson}
Hopkinson, G.R. , 1992, IEEE Trans Nucl Sci 39, 2018
\bibitem[2000]{Hornschemeier}
Hornschemeier, A., Brandt, W. N., Garmire, G .P. et al., 2000, ApJ 541, 49
\bibitem[2001]{Jansen}
Jansen, F., Lumb, D. H., Altieri, B. et al., 2001, A\&A 365, L1
\bibitem[2000]{Kuntz}
Kuntz, K. D. \& Snowden, S. L., 2000, Ap J 543, 195
\bibitem[2001]{Lehmann}
Lehmann, I., Hasinger, G., Schmidt, M., et al. 2001, A\&A, 371, 833
\bibitem[1988]{Lumb}
Lumb, D.H. \& Holland, A. D., 1988,  IEEE Trans Nucl Sci 35, 534
\bibitem[2002]{LUMBvig}
Lumb, D.H., Erd, C., Finoguenov, A. et al., 2002 in preparation  
\bibitem[Majerowicz \& Neumann 2001]{Majero}
Majerowicz, S. \& Neumann, D. M. in Galaxy Clusters and the High Red-shift Universe, eds D. Neumann, F. Durret \& J. Tran Thanh Van - Proceedings of XXIth Moriond Astrophysics Meeting (March 2001)
\bibitem[1980]{Marshall}
Marshall F. E., Boldt, E.A. and Holt, S.S. et al. 1980, Ap J 235, 4
\bibitem[1990]{McCammon} 
McCammon, D. \& Sanders, W. T. 1990, ARA\& A 28, 657
\bibitem[1985]{Mewe}
Mewe, R., Gronenschild, E.H.B.M. \& vd Oord, G. H. J., 1985, A\&AS 62, 197
\bibitem[2000]{Mushotzky}
 Mushotzky, R., Cowie, L., Barger, A. \& Arnaud, K. 2000, Nature 404, 459 
\bibitem[1993]{Plucinsky}
Plucinsky, P., Snowden, S., Briel, U. et al., 1993, Ap J 418, 519
\bibitem[2001]{Sabol}
Sabol, E. available at {\em http://heasarc.gsfc.nasa.gov/cgi-bin/Tools/xraybg/xraybg.pl}
\bibitem[2001]{Struder}
Str\"uder, L., Briel, U., Dennerl, K. et al., 2001, A\&A 365, L18
\bibitem[2001]{Tozzi}
Tozzi, P. , Rosati, P., Nonino, M. et al. 2001, Ap J 562, 42
\bibitem[2001]{Turner}
Turner, M. J. L. T., Abbey, A., Arnaud, M. et al., 2001, A\&A 365, L27
\bibitem[1997]{vanSpeybroek}
van Speybroek, L. P., Jerius,D., Edgar, R. J. et al., 1997, Proc SPIE 3113, 89
\bibitem[1999]{Ueda}
Ueda, Y., Takahashi, T., Inoue, H. et al., 1999, ApJ 518 656
\bibitem[1999]{Vecchi}
Vecchi, A., Molendi, S., Guainazzi, M. et al., 1999 A\&A 349, L73
\bibitem[2001]{Watson}
Watson, M. G., Augeres, J-L., Ballet, J. et al., 2001, A\&A 365 L51
\bibitem[2002]{Weisskopf}
Weisskopf, M. C., Brinkman, B., Canizares, C. et al., 2002 PASP 14 iss 791, 1
\end{thebibliography}
\end{document}